\documentclass[12pt]{article}

\usepackage{amsmath,amssymb,graphicx,multicol} 
\usepackage{epsf,color}
\usepackage{pstricks}
\usepackage{cite}

\newcommand{\beq}{\begin{eqnarray}}
\newcommand{\eeq}{\end{eqnarray}}

\newcommand{\SU}{{\rm SU}}
\newcommand{\U}{{\rm U}}
\newcommand{\centeron}[2]{{\setbox0=\hbox{#1}\setbox1=\hbox{#2}\ifdim

\wd1>\wd0\kern.5\wd1\kern-.5\wd0\fi
\copy0

\kern-.5\wd0\kern-.5\wd1\copy1\ifdim\wd0>\wd1
                                       \kern.5\wd0\kern-.5\wd1\fi}}
\newcommand{\ltap}{\>\centeron{\raise.35ex\hbox{$<$}}
                               {\lower.65ex\hbox{$\sim$}}\>}
\newcommand{\gtap}{\>\centeron{\raise.35ex\hbox{$>$}}
                               {\lower.65ex\hbox{$\sim$}}\>}

\newcommand\ZZ{\hbox{\zfont Z\kern-.4emZ}}
\font\zfont = cmss10 

\newcommand{\eq}[1]{(\ref{eq:#1})}

\begin{document}
\begin{titlepage}
\begin{flushright}
{\tt hep-ph/0611358}
\end{flushright}

\vskip.5cm
\begin{center}
{\huge \bf
The Gaugephobic Higgs}

\vskip.1cm
\end{center}
\vskip0.2cm

\begin{center}
{\bf
{Giacomo Cacciapaglia}$^{a}$,  {Csaba Cs\'aki}$^{b}$,
{Guido Marandella}$^{a}$, {\rm and}
{John Terning}$^{a}$}
\end{center}
\vskip 8pt

\begin{center}
$^{a}$ {\it
Department of Physics, University of California, Davis, CA
95616.} \\
\vspace*{.5cm}
$^{b}$ {\it Institute for High Energy Phenomenology\\
Newman Laboratory of Elementary Particle Physics\\
Cornell University, Ithaca, NY 14853, USA } \\
\vspace*{0.3cm}
{\tt  cacciapa@physics.ucdavis.edu, csaki@lepp.cornell.edu, maran@physics.ucdavis.edu, terning@physics.ucdavis.edu}
\end{center}

\vglue 0.3truecm

\begin{abstract}
\vskip 3pt \noindent We present a class of models that contains
Randall-Sundrum and Higgsless models as limiting cases. Over
a wide range of the parameter space $WW$ scattering is mainly
unitarized by Kaluza-Klein partners of the $W$ and $Z$, and the
Higgs particle has suppressed couplings to the gauge bosons. Such a 
gaugephobic Higgs can be significantly lighter than the 114 GeV LEP
bound for a standard Higgs, or heavier than the theoretical upper bound. These models predict a suppressed single top
production rate and unconventional Higgs phenomenology at the LHC:
the Higgs production rates will be suppressed and the Higgs
branching fractions modified. However, the more difficult the Higgs
search at the LHC is,  the easier the
search for other light resonances (like $Z', W', t'$, exotic
fermions) will be.

\end{abstract}

\end{titlepage}

\newpage


\section{Introduction}
\label{sec:intro}
\setcounter{equation}{0}
\setcounter{footnote}{0}
Extra-dimensions provide a natural explanation of the hierarchy
between the Planck and  the electroweak scale. In Randall-Sundrum
(RS) scenarios \cite{RS} the hierarchy is explained through an
exponential warping of mass scales along an extra-dimension.  Early versions of these
scenarios had all the Standard Model (SM) fields localized on the TeV
brane \cite{RS}. Later the gauge fields migrated to the
bulk~\cite{bulkgauge} and were soon followed by the quarks and
leptons \cite{matthias,GherPom}. This migration made it easier to
understand how the scenario could be  compatible with
unification and precision electroweak measurements even with TeV
scale  Kaluza-Klein (KK) modes for these fields.  However the Higgs remained
mainly\footnote{See however
refs.~\cite{LutyOkui,Davoudiasl:2005uu}.} on the brane, since this
was the main motivation of the RS scenario: to explain how the Higgs
could naturally have a TeV scale mass. However from the perspective
of the anti-de Sitter/conformal field theory (AdS/CFT)
correspondence, the localization of the Higgs on the TeV brane
corresponds to the assumption that the operator that breaks the
electroweak  symmetry has an infinite scaling dimension. Furthermore
from the AdS/CFT perspective the technical  naturalness of the Higgs
mass only requires that the operator have a scaling dimension
greater than two so that the operator corresponding to the Higgs
squared mass term has a dimension greater than  four and is thus
irrelevant. Therefore  it is interesting to explore what
happens in RS scenarios when the assumption of strict localization
of the Higgs field is relaxed.

Another generalization of RS scenarios was the suggestion
\cite{CGPT} that with an extra dimension, some or all of the $W$ and
$Z$ masses could come from the momentum along the extra dimension,
that is from the fact that their wave functions are not exactly
flat. By {\em increasing} the Higgs VEV on the brane the lightest
$W$ and $Z$ modes are forced to move further from the brane, and
their wave functions become less zero-mode like. In the limit that
the brane Higgs VEV goes to infinity the $W$ and $Z$ have no
couplings to the Higgs and all of their masses come from the
derivative of their extra dimensional profiles.  This is the
Higgsless limit \cite{CGPT}. One can then interpolate between the
Higgsless model, the conventional RS model and the SM by varying the
Higgs VEV and the Higgs localization parameter, and as a result  also
varying the size of the coupling of the Higgs to the gauge bosons.
Since the integral of the Higgs VEV contributes to the gauge
boson masses we cannot take the infinite VEV limit unless the Higgs
is strictly localized on the brane. Thus these bulk Higgs theories
generically have a physical Higgs with suppressed gauge
couplings. In conventional  RS scenarios the Higgs VEV is tuned to be
much smaller than the typical scale of the model (i.e. the scale of
the KK modes). Again from the AdS/CFT perspective this is not the
natural expectation. Normally we would expect at most a factor of $4
\pi$ between the analogs of the pion decay constant and the rho
mass.\footnote{A natural way of keeping the Higgs VEV below the KK
scale is via the pseudo-Goldstone mechanism. A concrete realization
of this idea in the RS context has been proposed in ref.~\cite{ACP}.} Of
course, at the LHC we would like to be prepared for the unexpected,
so when considering RS scenarios we should keep in mind that there
is a two dimensional parameter space (Higgs localization and VEV)
 to consider, and that the most ``natural" locations in
the parameter space are not those that have been studied so far. Due
to the repulsion of the wave functions of the bulk fields away from
the Higgs VEV we find that all Higgs couplings are generically
suppressed relative to the SM, and this has a variety of
consequences for experimental searches.

In this paper we will attempt a first survey of this gaugephobic Higgs
parameter space. In Section \ref{sec:model} we describe the class of models
(with many of the details relegated to the three appendices) and how
they are consistent with current experiments, while in Section \ref{sec:pheno}
we describe the phenomenology for upcoming experiments. To keep the
discussion simple we focus on two benchmark points: the first where
the Higgs couplings are about half of the SM values, and a second
where the couplings are about a tenth of  the SM values.  The Higgs
is not only harder to find, but it can also be outside the range of masses
allowed in the SM: it can be lighter than the experimental bound
from LEP  or heavier than the theoretical upper bound coming from
unitarity. Fortunately there are a variety of other particles that
can be searched for but typically it will take much longer for the
LHC to sort out the true situation than in other more simplistic
scenarios.

\section{The model}
\label{sec:model}
\setcounter{equation}{0}
\setcounter{footnote}{0}

In this section we examine  the effect of a bulk Higgs in an RS set-up using the
conformally flat metric
\beq ds^2=  \left( \frac{R}{z}\right)^2   \Big( \eta_{\mu \nu}
dx^\mu dx^\nu - dz^2 \Big)~, \eeq for $R<z<R^\prime$.  As usual the
UV brane is located at $z=R$ and the TeV brane is at $z=R^\prime$.
Here we will simply adapt the Higgsless model of ref.~\cite{CGPT,custodian} by
replacing the Dirichlet IR boundary conditions (BC's) (corresponding
to a very  large Higgs VEV localized on the IR brane) with a bulk
Higgs VEV. Let us first summarize the main features of the model:
the gauge group is SU(2)$_L \times$ SU(2)$_R\times$ U(1)$_{X}$, where
for the first two generations the $X$ charge is equivalent to $B-L$.
The gauge bosons for each group are denoted by $A^a_{L,R}$ and $B$,
with 5D gauge couplings $g_5$ and $\tilde g_5$ (for simplicity we
will take the two SU(2)'s to have the same gauge coupling). The
Higgs field is a bidoublet of $\SU(2)_L \times \SU(2)_R$:
\begin{equation}
\mathcal{H} = \left( \begin{array}{cc}
\phi_0^* & \phi_+ \\
- \phi_+^* & \phi_0
\end{array} \right) \qquad \mathcal{H} \rightarrow U_L \mathcal{H} U_R^\dagger\,;
\end{equation}
and its $\U(1)_{X}$ charge  is zero.
The Lagrangian  for the Higgs and the $\SU(2)_L\times \SU(2)_R\times \U(1)_{X}$ gauge sector is:
\beq\label{eq:gaugelag}
\mathcal{L} &=& \int_R^{R'}\, dz\, \frac{R}{z} \left\{ - \frac{1}{4 g_5^2} F_L^{aM N}F^{a}_{LM N}
- \frac{1}{4 g_5^2} F_R^{aM N} F^a_{R M N}
- \frac{1}{4 {\tilde g}_5^2} B^{M N}B_{M N}\right\} \\
&&+\int_R^{R'}\, dz\,  \left( \frac{R}{z} \right)^3 \left[\,{\rm Tr}
\left| \mathcal{D}_M \mathcal{H} \right|^2 - \frac{\mu^2}{z^2}\,{\rm
Tr} \left| \mathcal{H} \right|^2 \right] - V_{{\rm UV}}(\mathcal{H})
\delta (z-R) - V_{\rm TeV}(\mathcal{H})\delta (z-R') \,, \nonumber
\eeq where $\mu$ is a bulk mass for the Higgs (in units of the
inverse curvature radius $R^{-1}$). The potentials $V_{{\rm UV}}$ and $V_{\rm
TeV}$ on the branes determine the boundary  conditions for the Higgs
and  induce electroweak symmetry breaking (EWSB). In particular the
potential on the TeV brane which breaks the electroweak symmetry has
the usual form:
\begin{equation} \label{eq:Vtev}
V_{\rm TeV} = \left( \frac{R}{R'} \right)^4 \frac{\lambda R^2}{2} \left(  {\rm Tr}  \left| \mathcal{H} \right|^2 - \frac{v_{\rm TeV}^2}{2} \right)^2\,.
\end{equation}

The bulk equations of motion allow two solutions for the profile of
the bulk Higgs VEV (see appendix \ref{app:higgs}). We choose a UV
boundary condition that ensures that the allowed solution is
localized close to the TeV brane (i.e. electroweak symmetry is
broken at the TeV scale). The solution for the bulk profile can be
written as: 
\beq v (z) = \sqrt{\frac{2 (1+\beta) \log
R'/R}{(1-(R/R')^{2+2\beta})}}\, \frac{g V}{g_5} \frac{R'}{R} \left(
\frac{z}{R'} \right)^{2+\beta}\,, 
\label{eq:v(z)}
\eeq 
where $g$ is the SM $\SU(2)$
gauge coupling, and \beq \beta = \sqrt{4+\mu^2} \eeq is the
parameter determining how close to the TeV brane the Higgs VEV is
localized. In the presence of this VEV the gauge boson masses have
two sources: the curvature of their wave functions (also induced by
the VEV) and the direct overlap with the VEV. Instead of using
$v_{{\rm TeV}}$ of Eq. \eq{Vtev}, we choose to normalize the VEV
through the input parameter $V$ (appearing in Eq.~(\ref{eq:v(z)}), which carries the usual dimension
of mass). The factors of $\beta$, $R$ and $R'$ in front are chosen
in such a way that in the limit $V =  v_{{\rm SM}} \sim 246$ GeV the
direct contribution of the VEV to the gauge bosons mass saturates
the SM value, and the volume of the extra dimension shrinks to zero.
In other words one recovers the SM in 4 dimensions, independently of
the localization of the VEV.

The theory also contains a bulk physical Higgs boson. The mass of
the (possibly)  light mode is a free parameter and determined by the
quartic coupling $\lambda$ of the TeV  potential $V_{{\rm TeV}}$.
Similarly to the SM the mass of the physical Higgs is given by
\begin{equation}
  \label{eq:higgsmass2}
  m_h^2 \propto \lambda \; V^2\,.
\end{equation}
 It is clear  that increasing $V$ the Higgs can be
decoupled safely from the theory without entering a strongly
coupled regime in the Higgs sector, as  happens in the SM.

\begin{figure}[t]
  \centering
  \includegraphics[width=0.4\textwidth]{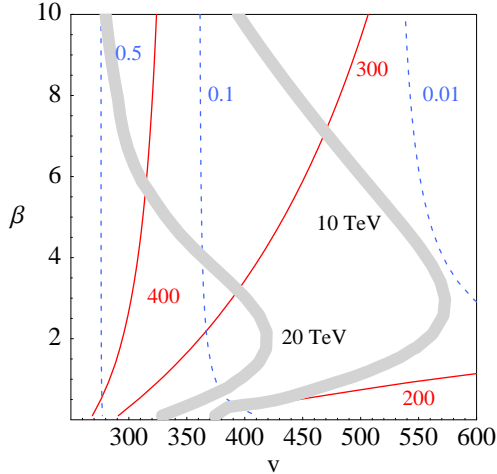}
  \caption{Lines with fixed values (400 Gev, 300 GeV, and 200 GeV) of the inverse size of the extra-dimension $R'$ in the $(V,\beta)$ plane (continuous lines), lines of fixed contribution of the Higgs to $WW$ scattering compared to the SM (dashed lines), and  lines of fixed cut-off of the theory due to the top sector becoming non-perturbative (thick gray lines). }
  \label{fig:radius}
\end{figure}

The model has 6 free parameters in the gauge-Higgs sector: $R$,
$R'$, the gauge couplings $g_5$ and $\tilde g_5$, the VEV parameters
$\beta$ and $V$. Three of the parameters ($R'$, $g_5$ and $\tilde g_5$)
can be determined by fixing the values of the $W$ and $Z$ mass, and
the coupling of the photon to the fermions. If we also fix
$1/R=10^8$ GeV~\footnote{Different values of $R$ will not
qualitatively affect our results.  The main effect of $R$ is to
enter in the relation between $M_W$ and $R'$: for instance, a smaller
$R$ will require a larger scale on the TeV brane
$1/R'$~\cite{CuringIlls}.}, we are left with the two parameters
$(V,\beta)$. In Fig.~\ref{fig:radius} we plot  contours of the IR
scale $1/R'$ as a function of $V$ and $\beta$. It is interesting to
notice various limits. First, when the VEV $V$ is equal to the SM
value $v_{{\rm SM}} \sim 246$ GeV, the IR scale becomes large. In
fact, in this case, the contribution of the Higgs VEV to the $W$
mass saturates the physical value of it, and thus the contribution
of the extra dimension has to vanish. As a consequence, in this
limit (with fixed $W$ mass), the volume of the extra dimensions has
to go to zero and the KK modes decouple: $R' \to R\,.$ In other
words, when $V \to v_{SM}$ we recover the 4D SM.

Another interesting limit is $\beta \to \infty$. This is a conventional
RS1 scenario \cite{ADMS}.  If one also takes the limit $V \gg 1/R'$
(i.e. top-right corner of the plot), one recovers the Higgsless
limit where the Higgs decouples  from the theory and can be made
arbitrarily heavy: the unitarization of the longitudinal $W$
scattering is then guaranteed by the contribution of the gauge boson
resonances~\cite{CGMPT,otherunitarity}. In the bulk of the parameter
space, however, we  find an intermediate situation where both the
Higgs and the gauge bosons KK modes contribute. In the SM, the Higgs
is forced to be lighter than about $1$ TeV in order to play a role
in the unitarization (unitarity bound). As a consequence, the SM
Higgs cannot escape detection at the LHC. In the Higgsless limit,
the Higgs cannot be detected in any experiment. In the intermediate
regime, the role of the Higgs in the unitarization can be less important
than in the SM, and the usual Higgs mass bound can be evaded. In
cases where the Higgs is out of reach of the LHC, the world will
appear to be Higgsless from the experimental point of view for
several decades. It is  important then to understand the unitarity bound
in this model. The Higgs enters in the longitudinal $W$ scattering
amplitude term which grows like the energy squared. The coefficient
of this term is proportional to

\begin{equation}
   \mathcal{A}^{(2)} \sim g_{WWWW}^2 - \frac{3}{4} \sum_{k} \frac{M_{Z^k}^2}{M_W^2} g_{WWZ^k}^2 - \frac{1}{4} \sum_{k} g_{WWH^k}^2\,,
\label{sumrule}
\end{equation}
where the sums cover all the KK modes. In the SM, only the first $Z$
and Higgs mode are present and both sums consist of only one term.
The presence of the gauge boson KK modes increases the second term,
allowing the third one (the Higgs  contribution) to be smaller. We
can thus define the following parameter to quantify how
``Higgsless'' the model is:

\begin{equation}
   \xi \equiv \frac{ \sum_{k} g_{WWH^k}^2}{g_{WWH}^2 (SM)} \,.
\end{equation}
In the SM limit $\xi \to 1$, while in the Higgsless limit $\xi \to
0$. Since 5D gauge invariance  guarantees the vanishing of
(\ref{sumrule}) this implies that the contribution of the lightest
Higgs will not be important until scales of order
$\Lambda_{SM}/\sqrt{\xi}$ where $\Lambda_{SM}$ is the unitarity
violation scale in the SM without a Higgs. This implies that the
Higgs mass can be raised to values approaching this number. In other
words, the unitarity bound on the Higgs mass can be relaxed by about
a  factor $\sim \sqrt{1/\xi}$: if the Higgs only contributes 10\%
with respect to the SM, it can be as heavy as about $3$ TeV; the
bound is raised up to about $10$ TeV if the contribution is only 1\%
($\xi=0.01$). We have plotted various contours for $\xi$ in
Fig.~\ref{fig:radius} (dashed lines).

It is well known that in Higgsless models particular care is needed
to maintain  compatibility with electroweak
precision tests (EWPT) while simultaneously getting a large enough  top quark mass.  Concerning EWPT, the main problem
turns out to be large corrections to the $S$ parameter. In order to
make them small, one has to allow the light fermions to be spread in
the bulk \cite{CuringIlls}. When their profile is approximately
flat, their wave function is orthogonal to the KK gauge bosons, and
the contributions to the $S$ parameter can be made arbitrarily
small. The top mass problem has been
successfully solved in ref. \cite{custodian}. One has to use an
alternative bulk-gauge/custodial representations for the third generation
\cite{CustodZbb}. Instead of having the left-handed (LH) top and bottom
transforming as a $({\bf 2,1})$ under $\SU(2)_L \times \SU(2)_R$ and
the right-handed (RH) fields as a  $({\bf 1,2})$, one has to embed
the LH top and bottom in a bidoublet  $({\bf 2,2})$, the RH top in a
singlet  $({\bf 1,1})$ and the RH bottom in a triplet  $({\bf
1,3})$. Localizing the LH bidoublet and the RH top close to the TeV
brane, while the RH bottom close to the Planck brane allows for
produce a large enough top mass without generating large deviations
to the $Z b_\ell \bar b_\ell$ coupling. More details are discussed
in Appendix \ref{app:fermion}. In the context of a bulk Higgs model exactly
the same methods  can be used to solve the two problems discussed
above. Of course as  $1/R'$ is increase to be larger than in the Higgsless limit, 
the bounds become less severe
and thus a larger region of parameter space is allowed.

Another potential issue is if the theory is perturbative in every
sector up to at least $\sim 5-10$ TeV, so  that higher order
corrections will not spoil EWPTs. We have to be concerned by the
gauge, top and Higgs sector. Using the techniques of Naive
Dimensional Analysis (NDA), properly adapted to 5D, one can estimate
the scale at which each sector becomes strongly coupled. The Higgs
sector, as long the lightest Higgs boson mode is lighter than about 1 TeV,
becomes strongly coupled at a very high scale. However, the gauge
and the top sector may still pose a potential problem. The strong
scale of the weak gauge coupling was estimated in~\cite{CuringIlls}:
it crucially depends on the IR scale $R'$. Even taking into account
factors as big as 4~\cite{Papucci}, we see that the strong scale is
safely above 10 TeV for $1/R' > 300$ GeV (a less conservative bound
of 5 TeV on the strong scale would imply $1/R' > 160$ GeV, however
this region is already unrealistic due to gauge boson resonances being too light). In
the top sector, we can estimate the strong scale corresponding to
the bulk top Yukawa, $y_5$, using NDA (properly warped down):

\begin{equation}
\Lambda_{top} \sim \frac{24 \pi^3}{y_5^2}\, \frac{R}{R'}\,,
\end{equation}
where we are neglecting unknown factors of order 1. The situation,
however, is more tricky, since the precise value of $y_5$ crucially
depends on the  localization of the left and right-handed tops. In
Fig.~\ref{fig:radius} we show two thick gray lines corresponding to
the top Yukawa becoming strong at 10 and 20 TeV, for $c_L = c_R = 0$
(this is in the range of parameters we use to produce a heavy
top mass and small corrections to the $Zb_\ell\bar{b}_\ell$ coupling). It is
interesting to notice that the more Higgsless the theory is, the
lower the strong scale is. This is a consequence of a lower IR
scale, and a smaller overlap with the top quark when we localize the Higgs on the
brane. However, in the strict Higgsless limit $V\to \infty$ this
scale becomes large again, due to the fact that in this limit
the Yukawa coupling goes to zero. In this limit, the cut-off of the
theory is set by the scale at which the gauge sector becomes strong.
We stress again that the value of this scale strongly depends
on the localization of the top quark. In the following we will be
conservative, and consider only points where the non-perturbative scale is above 20 TeV.

\section{Phenomenology}
\label{sec:pheno}

\begin{figure}[t]
  \centering
  \includegraphics[width=0.6\textwidth]{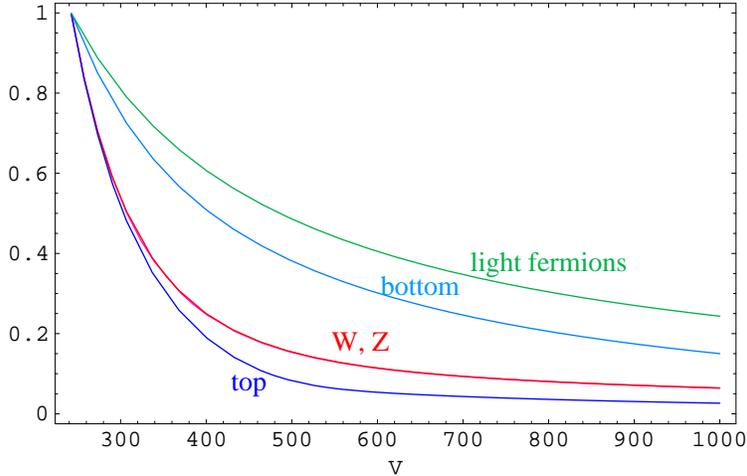}
  \caption{Suppression of the couplings of various SM fields with the Higgs with respect to the SM values for $\beta = 2$, as a function of $V$.}
  \label{fig:suppress}
\end{figure}

As in many other models with extra dimensions, bulk Higgs models predict
the existence of  KK states for all the SM fields. However, the most
distinctive feature is the presence of a Higgs boson with an
unconventional phenomenology.
In the region where $V \simeq 246$ GeV the  model approaches
the SM, and  the Higgs phenomenology is conventional. 
In the ``Higgsless'' limit ($V\gg 1/R', \beta\gg 1$) the Higgs
decouples from the theory: even if we want to make it light, through
a tiny quartic coupling \eq{higgsmass2}, its couplings to other
fields go to zero, and thus we will never be able to produce and
observe it. It is interesting to study some intermediate points,
where the couplings of the Higgs to the SM fields, although
suppressed, are large enough to lead to a potential discovery at the
LHC. In Fig.~\ref{fig:radius} we plotted the suppression of
the coupling of the Higgs to gauge bosons. In
Fig.~\ref{fig:suppress} we show the suppression of the couplings
with different SM fields as a function of $V$ (for fixed $\beta =
2$). The first thing to notice is that the Higgs couplings are more
suppressed to heavier particles. This can be understood in the
following way. The heavier the particle, the more it couples to the
VEV: this also implies that the wave functions are more distorted
away from zero-modes (they are more repelled from the
peak of the VEV). Schematically, the masses receive one contribution
from the direct overlap with the VEV, and the other from  terms involving the derivative along the extra
dimension (this contribution is still induced by the VEV, as the
light states would be zero modes for vanishing VEV). Thus, a larger
part of their mass comes from the extra dimension compared to
lighter particles. However, the (light) Higgs wave function is
essentially identical to the VEV profile, so that the coupling is
proportional to the portion of the mass coming directly from the
VEV. As we already discussed, this portion is larger for light
fermions compared to heavy ones. We can therefore see that the
couplings of the Higgs to heavy particles, like the top or weak gauge
bosons, are more suppressed than to light fermions, like the bottom.
This can affect the decay modes in an important way. Another
interesting point is that the production of the Higgs (which mostly
happens via the couplings to the $W$, $Z$ and $t$) is approximatively
suppressed  by the ``Higgslessness'' of the model: the parameter
$\xi$, plotted in Fig.~\ref{fig:radius}. In other words, in a point
where $\xi = 0.01$, the production rate of the Higgs is about 1\% of the
SM one.

To be more concrete, we choose two benchmark points: a) $V=300$ GeV
and $\beta=2$, b)  $V=500$ GeV and $\beta=2$. We report in
Table~\ref{tab:bench1} the couplings of the Higgs to various SM
fields, compared to the SM, in the two cases, assuming a Higgs mass of
120 GeV (these numbers have a mild dependence on the Higgs mass below 1 TeV).

\begin{table}[t]
  \centering
  \begin{tabular}{l|l}
    \multicolumn{2}{c}{a) $V=300 \; {\rm GeV},\; \beta=2$} \\
    \hline \hline
    $g_{ttH}$/SM & 0.52 \\
    $g_{WWH}$/SM & 0.54 \\
    $g_{ZZH}$/SM & 0.54 \\
    $g_{bbH}$/SM & 0.75 \\
    $g_{ffH}$/SM & 0.81 \\
  \end{tabular} \hspace{2cm}
  \begin{tabular}{l|l}
    \multicolumn{2}{c}{b) $V=500 \; {\rm GeV},\; \beta=2$} \\
    \hline \hline
    $g_{ttH}$/SM & 0.08 \\
    $g_{WWH}$/SM & 0.15 \\
    $g_{ZZH}$/SM & 0.15 \\
    $g_{bbH}$/SM & 0.38 \\
    $g_{ffH}$/SM & 0.49 \\
  \end{tabular}
  \caption{Higgs couplings to the SM fields for the two benchmark points a) $V=300$ GeV and  $\beta=2$, b) $V=500$ GeV and $\beta=2$. ``$f$'' stands for the light fermions }
  \label{tab:bench1}
\end{table}

First of all, the suppressed $HZZ$ coupling relaxes the LEP bound
$m_H>114$ GeV \cite{LEPHiggs}.  It turns out that for point a) the
production cross section is suppressed by a factor around 4, and
thus the bound becomes $m_H \gtrsim 95$ GeV  \cite{LEPHiggs}. For
point b) the production rate is suppressed by a factor of about 100,
so there is no limit from LEP. At the LHC, the suppressed couplings to
the top and the weak gauge bosons lead to a reduced production cross
section. Both gluon fusion and weak boson fusion production channels
receive a suppression of about a factor 4 in point a) and 100 in
point b). Concerning the decay modes, branching fractions to light particles are
enhanced with respect to the SM.


\begin{figure}[tb]
  \centering
  \includegraphics[width=0.5\textwidth]{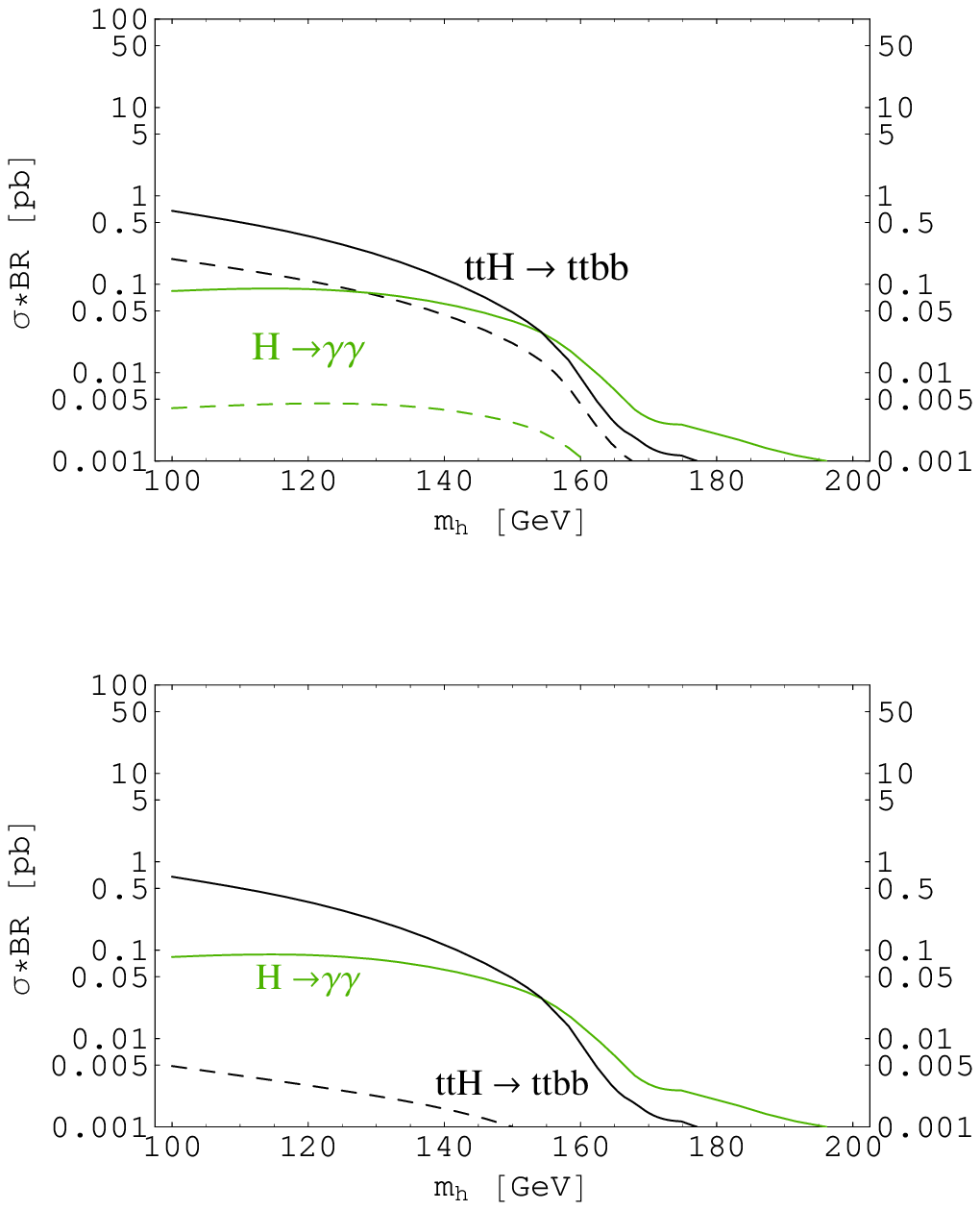}\includegraphics[width=0.5\textwidth]{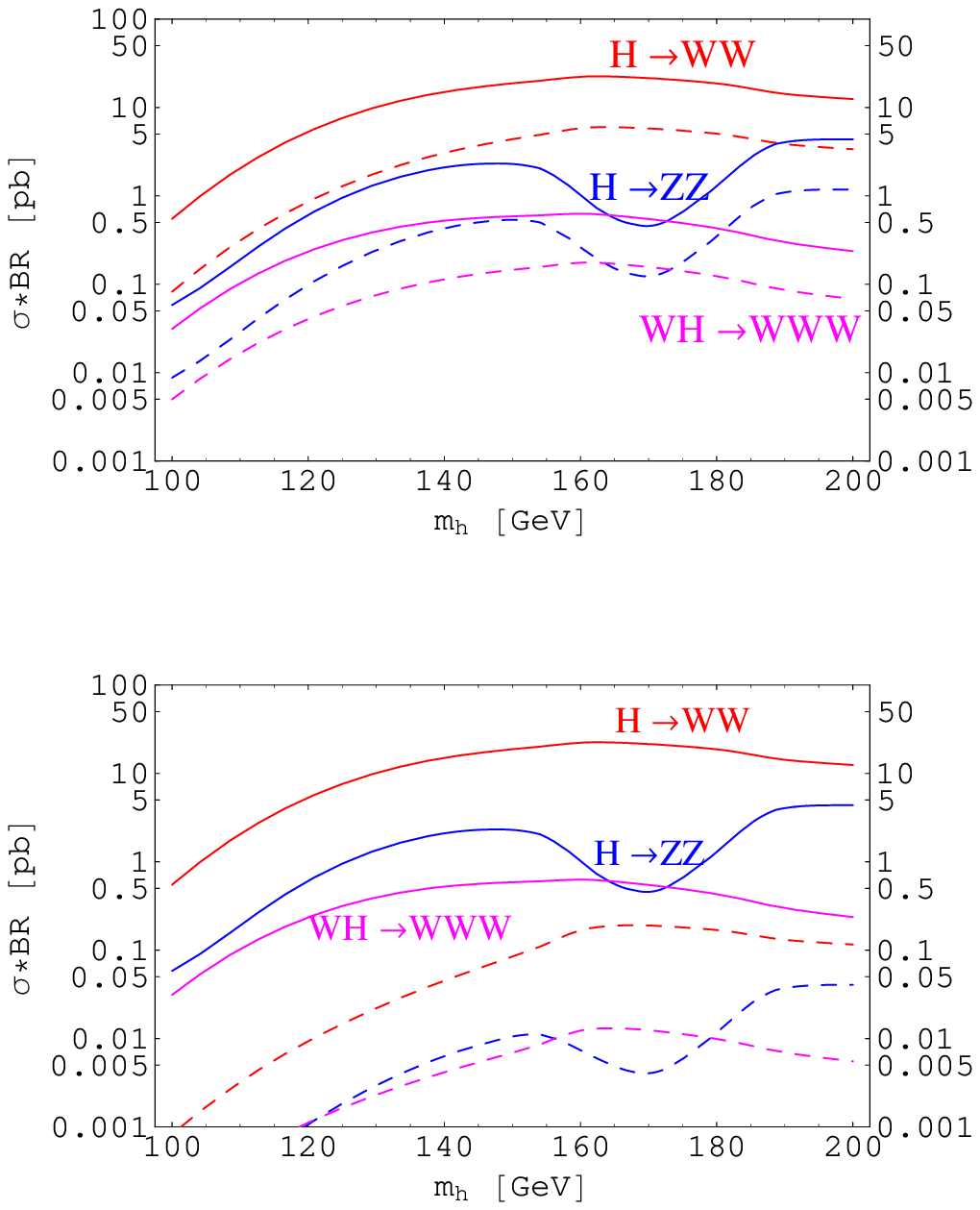}
  \centering
  \caption{Cross sections times branching ratios for various Higgs production and decay channels for the SM (solid lines) and gaugephobic Higgs (dashed lines) for $\beta =2$ with $V=300$ (top) and $V=500$ (bottom). }
  \label{fig:crossections}
\end{figure}

We show in Fig. \ref{fig:crossections} the production cross section
times the branching  fraction as a function of the Higgs mass for
a single Higgs decaying to $WW^{(*)},ZZ^{(*)},\gamma \gamma$ and
associated production $t \bar t H \to t \bar t b \bar b, \; WH \to
WWW^{(*)}$. These are the most promising channels for the discovery
at the LHC in various mass ranges. Continuous lines represent the SM,
while dashed lines represent the gaugephobic Higgs model. It is important to notice the Higgs
into two photons decay channel. It is mediated by a loop of tops and
$W$'s, which turn out to have the strongest suppression. For low
Higgs masses (when the $ZZ^*$ channel is still indistinguishable
from $Z\gamma^*$ background) in the SM there are two possible
discovery channels: $H \to \gamma \gamma$ and associated production
$t \bar t H \to t \bar t b \bar b$, with a similar discovery
significance. Since the coupling of the Higgs to $t$ and $W$ is more
suppressed than to the $b$, the $H \to \gamma \gamma$ channel is no
longer useful. All other channels get rescaled approximately by a
factor 4 in point a) and 100 in point b). Since the discovery
significance scales with the square root of the integrated
luminosity, it means that at the LHC it will be necessary to collect
16 and $10^4$ times the luminosity necessary in the SM case to claim
a discovery with the same statistical significance. Let us discuss a
concrete example. In the SM, discovering a $100$ GeV Higgs in the $t
\bar t H \to t \bar t b \bar b$ channel at a $\sim 8 \sigma$
significance requires an integrated luminosity of about 100 ${\rm
fb}^{-1}$ \cite{Gianotti}. In this model, in case a), and with the same
integrated luminosity, since the production rate decreases by  a
factor of 4 the significance would only be $\sim 2
\sigma$. A $5 \sigma$ discovery would require an integrated
luminosity of $100 \cdot (5/2)^2 \sim 600 \; {\rm fb}^{-1}$, which
almost saturates the total integrated luminosity ($700 \; {\rm
fb}^{-1}$) that the LHC is expected to collect. The discovery might
still be claimed, but the signal would be  delayed by
years and become statistically significant only at the very end
of the LHC run. Point b) is much more problematic: the
suppression factors are around 100, making the discovery of the
Higgs at the LHC impossible.

\begin{table}[t]
  \centering
  \begin{tabular}{l|l}
    \multicolumn{2}{c}{a) $V=300 \; {\rm GeV},\; \beta=2$} \\
    \hline \hline
    $ 1/R'$ & 372.5 GeV \\
    $W'$ &  918 GeV \\
    $Z'_1$ & 912 GeV \\
    $Z'_2$ & 945 GeV \\
    $G'$ & 945 GeV \\
  \end{tabular} \hspace{2cm}
  \begin{tabular}{l|l}
    \multicolumn{2}{c}{b) $V=500 \; {\rm GeV},\; \beta=2$} \\
    \hline \hline
    $ 1/R'$ & 244 GeV \\
    $W'$ &  602 GeV \\
    $Z'_1$ & 598 GeV \\
    $Z'_2$ & 617 GeV \\
    $G'$ & 617 GeV \\  \end{tabular}
  \caption{Spectrum of the first gauge boson KK states for the two benchmark points a) $V=300$  GeV and $\beta=2$, b) $V=500$ GeV and $\beta=2$.}
  \label{tab:KKGB}
\end{table}

As we discussed, there are regions in the parameter space where the
LHC can discover the Higgs  (even though requiring much more time
than the SM Higgs), and regions where the Higgs is definitely out of
reach for the LHC. It is therefore interesting, and complementary,
to look for other new particles in the gauge sector: the gauge boson
KK modes. As in the Higgsless model, they play a fundamental role in
the gauge sector, restoring the perturbative unitarity in the
longitudinal $W$ and $Z$ scattering. As a consequence, the ``more
Higgsless''  the models is, the lighter those resonances are. In
Table~\ref{tab:KKGB}, we show the spectrum of the lightest modes for
the two benchmark points. The phenomenology of these states was
already studied in~\cite{BMP}, where the authors only considered the
couplings among gauge bosons. They showed that the signal should be
observable for masses below $\sim 1$ TeV. Their approximation is
valid as long as the couplings with fermions is small: this is true
for light fermions~\cite{CuringIlls}, but not for the third
generation of quarks. The reason is that top and bottom are
localized towards the TeV brane, so that their wave functions have a
larger overlap with the KK states (localized on the TeV brane as
well). The study of vector resonances coupling negligibly to the
light fermions but sizeably to the third generation of quarks has
been performed in~\cite{Han:2004zh}. They show that, with an
integrated luminosity of 300 fb$^{-1}$,  a vector resonance can be
discovered at 5$\sigma$ significance up to about 2 TeV. The most
promising channels are $gg \to b \bar b, \; t \bar t$, with a $Z'$
radiated by one of the heavy quarks. The $Z'$ would then decay
mainly to top pairs, giving as a final state either $b \bar b t \bar
t$ or $t \bar t  t \bar t$. The choice of the couplings of the $Z'$
to the third generation made in~\cite{Han:2004zh}  is different from
our case: they consider a $Z'$ coupling only to right-handed top and
bottom quarks. However the right handed couplings of the top and the
bottom used in ~\cite{Han:2004zh} approximately match the sum of
left and right handed couplings of the third generation fermions in
this model, so we expect their result to be approximately applicable
to our model as well. Thus, for both our benchmark points the
discovery of the $Z'$ should not be missed at the LHC. The $Z'$ can
also be produced as a resonance in a Drell-Yan process. The coupling
to the light quarks is suppressed, but is still non-zero. Compared to
the previous case, one has one particle less in the final state,
which can compensate the effect of the suppression. For this reason,
the cross section might be comparable to the $Z' t \bar t$ and $Z' b
\bar b$ associated production. This channel would correspond to the
search of a heavy resonance in the $t \bar t$ channel.

\begin{table}[t]
  \centering
  \begin{tabular}{l|r|r|c}
   charge & a) $V=300 \; {\rm GeV}$ & b) $V=500 \; {\rm GeV}$ &  \\
    \hline
\hline
   5/3 & 581 GeV & 382 GeV &  $X_L$\\
   2/3 & 643 GeV & 511 GeV &  $T_L$\\
\hline
   -1/3 & 1062 GeV & 712 GeV & $b_R$ \\
   2/3  & 1058 GeV & 693 GeV & $T_R$ \\
   5/3  & 1124 GeV & 832 GeV & $X_R$ \\
\hline
   2/3  & 1160 GeV & 831 GeV & $t_L - t_R$\\
   -1/3 & 1242 GeV & 917 GeV & $b_L$ \\
   2/3  & 1318 GeV & 1114 GeV & $t_L - t_R$
        \end{tabular}
  \caption{Spectrum of the first resonances in the third generation quark sector, for the  two benchmark points. Near the mass eigenvalues, we report the field where such eigenstate mostly live. The effect of the Yukawa couplings is indeed to mix the representations, but numerically it corresponds a small shift of the masses.}
  \label{tab:KKfermions}
\end{table}

In the fermion sector, an interesting feature of this model is the
structure of the  resonances of the third generation quarks. This
structure may be used to probe the new representations that minimize
the deviations in the $Z b_\ell {\bar b}_\ell$ coupling. In
table~\ref{tab:KKfermions} we list the first fermionic resonances (below about
1 TeV) for the two benchmark points, and for a particular choice of
the bulk masses ($c_L = c_R^t=0$ and $c_R^b = -0.79$) that is
compatible with current data. There are many particles that are light
enough to be produced at LHC. The Yukawa interactions mix all the
representations, however their effect is numerically small (order
10\% in the masses) so that each particle is an approximate
interaction eigenstate (this is more true for small $V$): this
allows to try to reconstruct the non-trivial bulk gauge representations. For
instance, for the benchmark point a), we can easily identify
a doublet ($X$, $T$) with charges $+5/3$ and $+2/3$ respectively at
$\sim 600$ GeV as the lightest particles. At $\sim 1050$ GeV we have
a triplet of states that correspond to the triplet containing $b_R$,
and around $\sim 1200$ GeV we have a doublet ($t_L$, $b_L$) and a
singlet $t_R$. The discovery of a fermion with charge $5/3$ alone
would be a strong indication that a bidoublet exists, but from the spectra in the table we can see that a more
detailed structure can actually be observed notwithstanding the
mixings induced by the Yukawa couplings. 
The LHC should be able to distinguish the light doublet, and maybe more details of the spectrum depending on the precision in the mass determination.
The heavy top will be pair
produced in gluon fusion, or singly produced in $Wb$ or $Zt$ fusion
($gq \to T \bar b q'$ and $gq \to T \bar t q$). The techniques for
the search of heavy partners of the top have been analyzed in the
context of Little Higgs models. If the mass is below 1 TeV, as in
our case, its discovery at the LHC cannot be
missed~\cite{Azuelos:2004dm}. The exotic quark $X$ with charge $5/3$
is more interesting, as it is a direct consequence of the new
realization of the custodial symmetry. Similarly to $T$, it can
either be pair produced through gluon fusion or singly produced in
the process $gq \to X \bar t q'$. The $X$ would then decay to a $tW^+$
pair. The first mechanism would give 4 $W'$s and 2 $b$-jets, while
the second one  3 $W'$s and 2 $b$-jets: for both processes the SM
background would be very small, and thus we expect that the
discovery at the LHC should be guaranteed. Due to its charge, in the decay chain there will be two
same-sign $W$'s, $X \to W^+ t \to W^+ W^+ b$: in the leptonic channel
they will lead to same-sign lepton pair events.

Another consequence of the new realization of the custodial symmetry in the bulk, is a
suppression  in the $W t_l b_l$ coupling, that can be parameterized
in terms of a suppression of an effective CKM element $V_{tb}$~\footnote{In the language of KK states, this reduced coupling is indeed the effect of the mixing of the top with heavy KK fermions via the Yukawa couplings.}. If
the assumption of the unitarity of the CKM matrix is relaxed, such
suppression is not ruled out by present experiments~\cite{Vtb}. The
only way to directly probe this coupling is to measure  single
top production at the Tevatron and/or the LHC. In the two benchmark
points, we find that the effective $V_{tb}$ is 0.88 for $V=300$ GeV
and 0.74 for $V=500$ GeV (to be compared with the SM value $V_{tb}
\sim 1$). At the Tevatron, this implies a suppression of $23\%$
in the benchmark point a) and $45\%$ in point b) in the t-channel production cross section (and associated
$Wt$ production). In the s-channel, where the top is produced in
association with a bottom quark via a virtual $W$, we should also
take into account the interference with the $W'$s: this might induce
an even larger suppression~\cite{intWp}. At the LHC, the measurement
of single top production in the t-channel will allow a direct
measurement of $V_{tb}$ at a 5\% level~\cite{Vtb}, therefore testing
directly this prediction of the model. The s-channel is more
challenging to measure: due to the presence of 2 b-jets in the final
state (one is from the top decay) it suffers from a large background
from $t \bar t$ production.

Finally, the spectrum of KK excitations also contains massive gluons
(the masses  are listed in Table~\ref{tab:KKGB}). It is not a very
specific signature of this model, as it does not play any role in
EWSB and it is only a consequence of putting the $\SU(3)_C$ gauge
group in the bulk. Its coupling to the third generation is enhanced
for the same reason why the $Z'$ coupling is: a significant overlap
near the TeV brane. Thus it can be radiated away from a heavy quark.
The signature is very similar to the $Z'$ analyzed
in~\cite{Han:2004zh} but with bigger couplings: now they are $\SU(3)_C$
couplings rather than $\SU(2)$. Thus the LHC is sensitive to even
higher masses. In this case  the discovery of
the KK gluon $G'$ at the LHC cannot be missed.

\section{Conclusions}
\label{sec:conclusions}

We have begun an exploration of the space of generalized RS
scenarios where the Higgs localization width and VEV are taken as
free parameters. Generically the Higgs is gaugephobic and
topphobic. Such a gaugephobic Higgs can be lighter than the SM
experimental bound or heavier than the SM theoretical bound. A
gaugephobic Higgs will be more difficult (or impossible) to find at
the LHC.  However other particles (e.g. gauge  boson resonances) are
lighter than in conventional RS scenarios and thus will be easier to
find. The $Z^\prime$ is a good example of a resonance that should be
easy to look for. There are also top quark resonances and a tower of
exotically charged fermions that is necessary for the implementation
of the new realization of custodial symmetry. Since the top quark
couplings are also reduced in gaugephobic Higgs scenarios, it is
possible that the first experimental signature that can be found
will be the suppression of single top production at the Tevatron.

\section*{Acknowledgements}
We thank Bob McElrath, Lian-Tao Wang for useful discussions and comments. We also 
G.C., G.M. and J.T. are supported by the US Department of
Energy grant DE-FG02-91ER40674.
The research of C.C. is supported in part by the DOE OJI grant DE-FG02-01ER41206 and
in part by the NSF grant PHY-0355005. We also thank the Aspen Center for Physics
for its hospitality while part of this project has been performed.

\appendix

\section{The Higgs sector}
\label{app:higgs}
\setcounter{equation}{0}
\setcounter{footnote}{0}

The equations of motion for the Higgs field (with 4D momentum
$p_\nu$) and the BC's that   arise from the variation of the action
corresponding to Eq.~\ref{eq:gaugelag} are:
\begin{eqnarray}
\left( z^3 \partial_z \frac{1}{z^3} \partial_z + p^2 - \frac{\mu^2}{z^2} \right) \mathcal{H}  & = & 0\,, \label{eq:eom}\\
\left. \left(\frac{R}{R'}\right)^3 \partial_z \mathcal{H} +  \frac{\partial}{\partial \mathcal{H}^*} V_{\rm TeV} \right|_{R'} & = & 0\,,\label{eq:BCtev}\\
\left. \partial_z \mathcal{H} - \frac{\partial}{\partial \mathcal{H}^*} V_{{\rm UV}} \right|_{R} & = & 0\,. \label{eq:BCplanck}
\end{eqnarray}
We will assume that the quartic term is localized on the IR brane, as we want EWSB to take place there.

In order to find the VEV we need to solve the equations of motion
and find a solution that  minimizes the full potential, as encoded
in the BC's. For $p_\nu=0$ and a diagonal VEV which breaks SU(2)$_L
\times$ SU(2)$_R \to$ SU(2)$_D$: \beq < \mathcal{H} > = \left(
\begin{array}{cc}
1 & 0 \\
0 & 1
\end{array} \right) \frac{v (z)}{\sqrt{2}}\,.
\eeq
 the solutions are of the form
\begin{equation}
v (z) = a \left( \frac{z}{R} \right)^{2+\beta} + b \left( \frac{z}{R} \right)^{2-\beta}\,,
\label{eq:vevsolution}
\end{equation}
where $\beta = \sqrt{4+\mu^2 }$.  Note that the dimension of the corresponding CFT operator is $2 \pm \beta$ depending on the UV boundary conditions.
We can see that the effect of $\beta$, i.e. of the bulk mass, is to control the localization of the profile of
the VEV  in the bulk.
The bulk mass squared can be negative but  it is bounded \cite{Breitenlohner} by $\mu^2 > - 4$.
The form of the localized potentials $V_{\rm TeV}, V_{{\rm UV}}$ determines if a VEV is generated and its size.
We want a VEV generated on the TeV brane, so we will add a ``mexican hat'' potential there.
On the UV brane, on the other hand, we just add a mass term:

\begin{equation} \label{eq:planckpot}
V_{{\rm UV}} = m_{{\rm UV}}\, {\rm Tr}  \left| \mathcal{H} \right|^2\,,
\end{equation}
where $m_{{\rm UV}}$ has dimension of mass.
Eq.~\ref{eq:BCplanck} can be used to fix the ratio between the two coefficients in the solution (\ref{eq:vevsolution}):
\begin{equation}
\frac{b}{a} = - \frac{2+\beta -m_{{\rm UV}} R}{2 - \beta - m_{{\rm UV}} R}\,.
\end{equation}
The relation between the  mass $m_{{\rm UV}}$ and the bulk mass can
select one of the two  solutions: $m_{{\rm UV}} R = 2 \pm \beta$
will select the solution growing towards the TeV or the UV brane
respectively. In the CFT language, this boundary condition
corresponds to the determining the CFT operator (and its scaling
dimension) represented by the bulk Higgs. In the following we will
assume that only the solution with the IR localized profile
(corresponding to the operator with the higher dimension) is
selected by the UV mass so that: \beq
 \label{eq:UVmass}
m_{{\rm UV}}& = &\frac{2+ \beta}{R}\\
 \label{eq:VEVprofile}
v (z) &=& a \left( \frac{z}{R} \right)^{2+\beta}~. \eeq Note that
the dimension of the corresponding CFT operator which breaks
electroweak symmetry is $2+ \beta$: being $\beta\ge0$, in this case the Higgs mass is naturally of order TeV ($1/R'$). 
Tuning the UV mass as in Eq.~\ref{eq:UVmass} is not essential: the
solution we select is in fact growing faster towards the TeV brane,
so it will have a larger effect on the $W$ and $Z$ mass. We will
keep only one solution just to simplify our calculation. Note also
that if we wanted to select the other solution ($z^{2-\beta}$), all
we would have to do is to flip the sign of $\beta$ in all the
equations: in other words, taking $\beta<0$ formally corresponds to
the other solution. This will only be interesting for one reason:
for $\beta = -1$ ($v \sim z$) we have a flat profile for the Higgs
VEV.

On the TeV brane, we add the following potential:
\begin{equation} \label{eq:tevpot}
V_{\rm TeV} = \left( \frac{R}{R'} \right)^4 \frac{\lambda R^2}{2} \left(  {\rm Tr}  \left| \mathcal{H} \right|^2 - \frac{v_{\rm TeV}^2}{2} \right)^2\,,
\end{equation}
where the warp factors have been added so that all the parameters
have a natural scale set by $R$. Note also that a factor of $R^2$
has been added to make $\lambda$ dimensionless, while $v_{\rm TeV}$
has  dimension [mass]$^{3/2}$. Imposing the BC in
Eq.~(\ref{eq:BCtev}), the coefficient $a$ in
Eq.~(\ref{eq:VEVprofile}) is fixed. We can thus write the profile of
the VEV  as
\begin{equation}
v(z) = \frac{1}{R^{3/2}}  \left(R^3 v_{\rm TeV}^2 - \frac{2 (2+\beta)}{\lambda} \right)^{1/2} \left(\frac{z}{R'} \right)^{2+\beta} \,.
\end{equation}
The VEV $v(z)$ is real, so  there is a solution only if $v_{\rm
TeV}$ is large enough: a TeV brane  localized negative squared mass
has to be big enough to overcome the effect of the positive bulk
mass. We can define a 4D VEV as the integral of the 5D VEV along the
extra-dimension:
\begin{equation} \label{eq:4DVEV}
v_4^2 =  \int_R^{R'}\, dz\, \left( \frac{R}{z} \right)^3 v^2 (z).
\end{equation}
In the limit where the VEV is localized on the IR branes, $v_4$ is the value of the 4D VEV.
Thus the 5D VEV can be rewritten as:
\begin{equation} \label{eq:5DVEV}
v (z) = \sqrt{\frac{2 (1+\beta)}{R^3 (1-(R/R')^{2+2\beta})}}\, v_4 R' \left( \frac{z}{R'} \right)^{2+\beta}\,.
\end{equation}
Notice that the natural size of $v_4$ is $\sim 1/R'$.

At this point we can parameterize the bulk Higgs through three free
parameters: the bulk mass $\beta$,  the 4D VEV $v_4$ and the quartic
coupling $\lambda$. The first controls the localization of the
profile, the second  sets the size of the Higgs VEV, while the third
controls the mass of the lightest physical Higgs mode (which we will
come back to later).

The only parameter that carries a dimension of mass is $v_4$.
In order to make its value more physically meaningful, we can rescale it, and
reparameterize the Higgs VEV $v$ as: 
\beq v (z) = \sqrt{\frac{2
(1+\beta) \log R'/R}{(1-(R/R')^{2+2\beta})}}\, \frac{g V}{g_5}
\frac{R'}{R} \left( \frac{z}{R'} \right)^{2+\beta}\,, \eeq 
where $g$
is the SM $\SU(2)$ gauge coupling, and $V$ is the new input parameter
we are defining. The reason for this choice is that in the flat VEV
case ($\beta \to -1$) the contribution of the VEV  in the equation
of motion for the gauge bosons \eq{eomALmAR} is 
\beq \frac{R^2}{z^2} \frac{g_5^2}{4} v^2
(z) \xrightarrow[\beta \to -1]{} \frac{g^2 V^2}{4}\,; \eeq 
Since for
$\beta \to -1$ the $W$ wave function is flat, the $W$ mass comes
entirely from the Higgs VEV. Thus in this limit $V$ has to be equal
to the SM VEV $v_{SM} \sim 246$ GeV. We numerically checked that we obtain the SM when $V \to v_{SM}$ independently of the localization of the VEV.

The bulk Lagrangian for the physical Higgs $h$ is:
\begin{equation}
\mathcal{L}_h = \int_R^{R'}\, dz \, \left( \frac{R}{z} \right)^3 \left\{ \frac{1}{2} \left( \partial_\mu h \right)^2 - \frac{1}{2} \left( \partial_z h + \partial_z v \right)^2 - \frac{1}{2} \frac{\mu^2}{z^2} \left( h + v \right)^2 \right\}~.
\end{equation}
The tadpoles in $h$ cancel out due to the equations of motion of the Higgs VEV $v(z)$.
On the other hand, the localized potentials do generate a mass term for the Higgs.
The equations of motion for a mode with $p_\nu p^\nu=m^2$ are:
\beq
\displaystyle \left( z^3 \partial_z \frac{1}{z^3} \partial_z + m^2 - \frac{\mu^2}{z^2} \right) h = 0\,,\nonumber \\
\left. \partial_z h - m_{{\rm UV}} \; h \right|_{R} = 0\,,\\
\left. \partial_z h + (R/R') \; m_{\rm TeV} \; h \right|_{R'} = 0\,.\nonumber
\eeq
The mass on the UV brane is given by Eq. (\ref{eq:UVmass}).
The solution of the bulk equation of motion and the UV brane boundary condition is given by
\begin{equation}
h_m (z) = A z^2 \left( Y_{1+\beta} (m R) J_\beta (m z) + J_{1+\beta} (m R) Y_\beta (m z) \right)~,
\end{equation}
where $A$ is a normalization factor. The mass eigenvalues $m$ are determined by the TeV localized potential.
On the TeV brane, the effective mass term is related to the Higgs VEV and the quartic coupling $\lambda$:
\begin{equation}
m_{\rm TeV} R = \lambda R^3 v^2 (R') - (2+\beta)\,.
\end{equation}
The value of this parameter determines the Higgs spectrum. As in the
SM, the Higgs mass is a free parameter, determined by $\lambda$.
However, $\lambda$ will also determine the Higgs quartic coupling,
so we need to make sure it is not too  large in order to avoid
strong coupling in the Higgs sector.

\section{Gauge sector} \label{app:gauge}
\setcounter{equation}{0}
\setcounter{footnote}{0}

We now analyze the gauge sector of this model. Expanding the bulk scalar around the VEV we have:

\begin{equation}
\mathcal{H} = \frac{1}{\sqrt{2}} \left( v(z) + h \right) \left( \begin{array}{cc}
1 & 0 \\
0 & 1
\end{array} \right) e^{i \pi^a \sigma^a}\,.
\end{equation}

The $\pi^a$s are the Goldstone bosons eaten by the broken gauge directions, associated with the combination of fields $A_L^a - A_R^a$.
We will focus on those gauge fields and Goldstones: to keep the formulae simple and clear, we will use the notation of a simple group completely broken by the VEV, and a generalization to our model will be straightforward.
The Lagrangian up to quartic terms becomes:

\begin{multline} \label{eq:lgauge}
\mathcal{L}_{gauge} =  \int_R^{R'}\, dz\, \frac{R}{z} \frac{1}{g_5^2} \left\{ - \frac{1}{4} F_{\mu \nu}^2 + \frac{1}{2} \left( \partial_z A_\mu \right)^2 + \frac{1}{2} \left( \frac{R}{z}\right)^2 g_5^2 v^2 A_\mu^2 + \right. \\
 + \frac{1}{2} \left( \partial_\mu A_5 \right)^2 + \frac{1}{2} \left( \frac{R}{z} \right)^2 g_5^2 v^2 \left( \partial_\mu \pi \right)^2 - \frac{1}{2} \left( \frac{R}{z} \right)^2 g_5^2 v^2 \left( \partial_z \pi + A_5 \right)^2 +\\
\left. -\left( z \partial_z \frac{A_5}{z} + \left( \frac{R}{z} \right)^2 g_5^2 v^2 \pi - A_5 \delta (z-R) + A_5 \delta (z-R') \right) \partial_\mu A^\mu \right\}\,,
\end{multline}
where $g_5$ is the 5D gauge coupling of the SU(2)'s.
Note that the Higgs VEV always appears in the following combination:

\begin{equation}
\tilde v (z) = \frac{R}{z} g_5 v(z) = \sqrt{\frac{2 (1+\beta)}{(1-(R/R')^{2+2\beta})}}\, \frac{g_5}{\sqrt{R}}\, v_4 \left( \frac{z}{R'} \right)^{1+\beta}\,.
\end{equation}
This quantity carries more physical meaning, as it is equivalent to
a bulk mass for the gauge  bosons. In general the gauge boson masses
have two sources: the curvature of their wave function and the Higgs
VEV. Notice also that if $\beta=-1$ the VEV is flat, the gauge boson
wave function is also flat, and the mass term becomes constant:

\begin{equation}
\tilde v (z) \rightarrow \frac{g_5}{\sqrt{R \log R'/R}}\, v_4 = g_4 v_4\,.
\end{equation}
In this case the only source for the gauge boson mass is the Higgs VEV.

The mixing terms on the third line of Eq.~\ref{eq:lgauge} can be
cancelled out by $R_\xi$ gauge fixing terms  in the bulk and on the
branes:

\begin{multline}
\mathcal{L}_{GF} =  \int_R^{R'} \, dz\, - \frac{R}{z} \frac{1}{g_5^2} \left\{ \frac{1}{2 \xi} \left( \partial_\mu A^\mu - \xi \left( z \partial_z \frac{A_5}{z} + \tilde v^2 \pi \right) \right)^2 + \right.\\
\left. + \frac{1}{2 \xi_{{\rm UV}}} \left( \partial_\mu A^\mu - \xi_{{\rm UV}} A_5 \right)^2 \delta (z-R) + \frac{1}{2 \xi_{{\rm TeV}}} \left( \partial_\mu A^\mu + \xi_{{\rm TeV}} A_5 \right)^2 \delta (z-R') \right\}\,.
\end{multline}
The unitary gauge is, as usual, given in the limit where all the three $\xi$-parameters are sent to infinity.
A combination of the $\pi$'s and $A_5$'s is eaten by massive gauge bosons, while another one gives rise to a tower of physical states. However, those scalars are as heavy as the KK states, and the do not play any interesting role in the symmetry breaking.
In the unitary gauge, the vector bosons obey the following equation of motion:
\begin{equation}
\left( z \partial_z \frac{1}{z} \partial + m^2 - \tilde v^2(z) \right) A_\mu (z) = 0.
\end{equation}

Ii is now straightforward to write the equations of motion for the gauge fields in our specific model:
\beq
\left( z \partial_z \frac{1}{z} \partial + m^2 - \frac{g_5^2}{4} \frac{R^2}{z^2} v^2 \right) (A_{L\mu}^a - A_{R\mu}^a) &=& 0\,, \label{eq:eomALmAR}\\
\left( z \partial_z \frac{1}{z} \partial + m^2 \right) ( A_{L\mu}^a + A_{R\mu}^a ) &=& 0\,, \\
\left( z \partial_z \frac{1}{z} \partial + m^2  \right) B_\mu &=& 0\,.
\eeq
As in the Higgsless model, the BC's on the UV brane break SU(2)$_R\times$U(1)$_{X} \to$U(1)$_Y$:
\beq
\begin{array}{cc}
\partial_5 A_{L\mu}^a = 0\,, &
 A_{R\mu}^\pm  =  0\,,\\
\partial_5 \left( \frac{1}{\tilde g_5^2} B_{\mu} + \frac{1}{g_5^2} A_{R\mu}^3 \right) = 0\,, & A_{R\mu}^3 - B_\mu  = 0\,.
\end{array}\eeq
On the TeV brane all the gauge fields have Neumann BC's since EWSB is accomplished by the Higgs VEV.

\section{The fermion sector} \label{app:fermion}
\setcounter{equation}{0}
\setcounter{footnote}{0}

The conventional choice for embedding fermions into $\SU(2)_L \times
\SU(2)_R$ models is to put the LH and RH SM fermions into $SU(2)_L$
and $SU(2)_R$ bulk doublets respectively. This is what we will be
using for the light fermions. For instance, $\Psi_L=(u_L, d_L)$,
$\Psi_R=(u_R, d_R)$ transforming as \beq \label{oldreps} ({\bf
2},{\bf 1})_{1/6},~\,\,\,\,\, ({\bf 1},{\bf 2})_{1/6} \eeq
 of $SU(2)_L \times
SU(2)_R \times U(1)_{X}$. For these representations, the $X$-charge
can be identified as $X=(B-L)/2$. The SM zero  modes
\cite{matthias,GherPom,CGHST} can be reproduced by the assignment of
the following BC's: \beq
\begin{array}{c|cc}
 \Psi_L & \mbox{UV} & \mbox{IR} \\
 \hline
\chi_L = \vphantom{ \sqrt{\left( \begin{array}{c} \chi_{u_R} \\ \chi_{d_R} \end{array} \right) }}
 \left( \begin{array}{c} \chi_{u_L} \\ \chi_{d_L} \end{array} \right) & + & + \\
\psi_L = \left( \begin{array}{c} \psi_{u_L} \\ \psi_{d_L} \end{array} \right)& - & -
\end{array}
 \qquad
\begin{array}{c|cc}
\Psi_R & \mbox{UV} & \mbox{IR} \\
\hline
\chi_R =  \vphantom{ \sqrt{\left( \begin{array}{c} \chi_{u_R} \\ \chi_{d_R} \end{array} \right) }}
\left( \begin{array}{c} \chi_{u_R} \\ \chi_{d_R} \end{array} \right) & - & -  \\
\psi_R = \left( \begin{array}{c} \psi_{u_R} \\ \psi_{d_R} \end{array} \right)  & + & +
\end{array}
\eeq where $+$ stands for a Neumann BC, $-$ stands for a
Dirichlet BC, $\chi$ represents the LH chirality, and $\psi$ the
RH chirality. To give the fermions a mass, a bulk Yukawa coupling can be written. After replacing the Higgs with its VEV the mass term has the form 
\beq \label{eq:fermionmass} \frac{\lambda}{\sqrt{2}} v(z) \left( \chi_L \psi_R
+ \chi_R \psi_L \right) + h.c. \eeq. Note that in order
to preserve custodial symmetry, the Yukawa coupling
(\ref{eq:fermionmass}) couples to both up and down type quarks. To
split their mass, a  large kinetic term can be added on the UV brane
for $\psi_{d_R}$, since $SU(2)_R$ is broken on that brane.

For the third generation, things are more problematic due to the
large top mass. In order to generate  the top mass, one needs to
localize it closer to the IR brane, where the Higgs lives. However
this turns out to generate large corrections to the $Z b_\ell \bar
b_\ell$ coupling. To solve this problem, one has to use a different
set of representations for the third generation, as proposed by
\cite{CustodZbb}. The LH doublet is now embedded in a bidoublet of
$SU(2)_L \times SU(2)_R$, the RH top in a singlet, and the RH bottom
in a triplet of $SU(2)_R$: \beq \Psi_L = ({\bf 2,2})_{2/3} = \left(
\begin{array}{cc}
t_L & X_L \\
b_L & T_L
\end{array} \right), \qquad
 \Psi_R = ({\bf 1,3})_{2/3} = \left( \begin{array}{c}
X_R\\ T_R \\ b_R
\end{array} \right), \qquad
t_R =({\bf 1,1})_{2/3}, \eeq where all these fermion fields are bulk
fields. As usual, $Y=T^3_{R}+X$,  $Q=T^3_{L}+Y$, and the BC's ensure
that the only zero modes correspond to  SM fields. There are two
SU(2)$_L\times$SU(2)$_R\times$U(1) invariant Yukawa couplings that one
can write in the bulk. After plugging in the VEV for the bidoublet
Higgs field these will lead to the bulk mass terms for the fermions
\beq \mathcal{L}_{Y} = \lambda_3 \frac{v(z)}{\sqrt{2}}
\left[\frac{1} {\sqrt{2}} T_R\, \left(t_L + T_L \right) +  b_R b_L +
X_R X_L\right] + \frac{\lambda_1 v(z)}{2} \; t_R\, \left(t_L -T_L
\right) +h.c. \eeq Under the unbroken $SU(2)_D$ subgroup, the
combination $(t_L -T_L)/\sqrt{2}$ is the singlet component of
$\Psi_L$, while the fields $(X_L,(t_L +T_L)/\sqrt{2},b_L)$ form the
triplet. Choosing $\Psi_L$ and $t_R$ to be localized close to  the
IR brane, the top mass can easily obtained. In order to avoid large
deviations to the $Z b_\ell \bar b_\ell$, one has also to localize
$\Psi_R$ close to the UV brane \cite{custodian}. Changing $V$ and
$\beta$ changes the allowed parameter space, which becomes larger
for larger values of $1/R'$.

\end{document}